# 基于夜光遥感对于区域和城市群的发展的研究

张炅焱

（武汉大学　遥感信息工程学院，湖北　武汉　430000）

**摘　要**：为了尝试从夜光遥感影像的角度对于区域经济发展和城市扩大化现象进行研究，研究者利用 NOAA 提供的夜光遥感影像数据(1992 年～2013 年数据)，使用 ArcGIS 等软件进行图像信息处理，获取了图像特定区域的基本像素信息数据，并从时空域上分析这些数据，以呈现出中国近些年来的区域经济发展态势，并且尝试探讨中国经济迅速发展所带来的城市化效应。经过对于数据的分析和研究，结果表明，中国城市化发展速度仍处于高峰时期，具有极大的发展潜力和发展空间。但与此同时，人们也需要注意到地区发展的不均衡问题。

**关键词**：夜光遥感；区域经济；城市发展；城市经济

**Abstract:** In order to study the phenomenon of regional economic development and urban expansion from the perspective of night-light remote sensing images, researchers use NOAA-provided night-light remote sensing image data (data from 1992 to 2013) along with ArcGIS software to process image information, obtain the basic pixel information data of specific areas of the image, and analyze these data from the space-time domain for presentation of the trend of regional economic development in China in recent years, and tries to explore the urbanization effect brought by the rapid development of China's economy. Through the analysis and study of the data, the results show that the urbanization development speed in China is still at its peak, and has great development potential and space. But at the same time, people also need to pay attention to the imbalance of regional development.

# 1 前言

自从改革开放以来，中国经济就以一种很快的速度进行发展。但是在空间视角下，这种发展是以什么样方式进行的，则很引人感兴趣。从 20 世纪 80 年代中期到 90 年代后期，中国开启了大规模扶贫政策，带动了部分贫困地区的发展[1]。二十一世纪来，中国的市场经济政策初见成效，中国在各个方面都取得长足的进步，研究者期望从一个新的角度来认识中国的变化。此时，夜光遥感技术为研究者提供了绝佳的研究工具和数据源。

随着经济的快速发展和城市化进程加速，夜光遥感可以更加快捷直观地反映人类活动的特点，使得研究者对于经济发展和人文活动的认识变得更加便捷、省力[2]。研究者所采用的夜光遥感产品是具有连续的 DN 值的灰度图像，而这种灰度图像所呈现出来的明暗程度可以让研究者很直观地感受到这个区域的繁荣程度。一般的，如果通过目视解译的话，研究者可以有如下的判断准则：即，当一个区域的亮度越高(即 DN 值越高)的时候，研究者就可以认为这个地方的经济也更加发达。

究其原因，研究者知道，随着经济社会的发展和进步以及夜间照明设备的普及，人类文明的标志性特征之一就是夜间的灯光。当夜空处于无云状态的时候，研究者可以利用遥感卫星捕捉到城镇灯光、渔船灯光、火点等可见光辐射源[3]。经过对于影像的一些处理，研究者就可以使用这些影像，以求对于人类的活动和发展做出评估。

为了获取人类夜间活动所引起的地球表面亮度分布状况，20 世纪 70 年代开始，美国开展了国防气象卫星计划(Defense Meteorological Satellite Program, DMSP)，且其中部分卫星装备了线性扫描业务系统(Operational Linescan System, OLS)，可以获取夜间地表微弱的可见光辐射，并且得到一系列年度的无云夜间灯光影像[4]。我们知道，DMSP/OLS 产品是一种持续时间长、年际波动较小以及经过空间清晰化处理的

卫星观测产品[5]。在这里，研究者正是利用 DMSP/OLS 产品来获取研究者所需要的信息的。

一些学者通过对于夜光遥感影像的研究发现，由人类活动直接引起的夜光辐射信号在区域尺度上与众多城市化和社会经济变量之间存在显著的定量关系，并且这种关系在时空域方面上通常呈现出单调和稳健的特点[6]。因此从夜光遥感影像的图像中，研究者可以对于其所划分的目标区块内的 DN 值的分布进行一些数理统计，计算出研究者所需要的一些数据，从而对于这些数据进行时空上的评估和可视化，然后得到研究者对于区域发展的分析结论。

这里的评估目标主要是集中在如下几点：(1)评价一个区域的经济增长速度；(2)评价一个区域的经济发展均衡程度；(3)评价一个地区的城市化程度和城市化发展。这些分析都是建立在地区经济实力(如 GDP 指标等)的发展和夜光遥感影像的 DN 值变化相匹配的情况下。经过一些学者的分析，研究者可以发现一些国家的 GDP 增量和夜光增长量之间的拟合函数回归系数是 0.9187，充分证明了夜光遥感可以反映出一个地区的经济实力，进而反馈出地区社会和经济的发展动态[7]。

目前，很多相关研究人员都通过各种方法来利用夜光遥感影像对于区域 GDP 进行计算或者是对城区范围进行数据统计，以求从定量的角度上对于夜光遥感影像进行分析和研究。本文也依照定量研究的原则，通过对于影像的数据统计，以对中国经济的阶段性历史发展进行阐释。同时，本文也会结合夜光遥感影像本身，来呈现出这些统计数据的实际意义，以求在更广的范围对于区域经济发展和城市化进度进行解释。

综上所述，笔者将利用夜光遥感 DN 值作为直接使用数据，从 DN 值来分析区域的经济增速、发展均衡度，并尝试从空间几何和具体数据中对于区域城市化进度进行描述和解释，以求从一个新的角度来理解经济发展和城市化特点，为决策者开辟新的参考范畴。

# 2 对于夜光遥感数据进行的处理和数据提取

## 2.1 数据预处理

对于数据进行预处理是十分重要的。事实上，DMSP/OLS 系列产品是由多种传感器所得到的，具体有 F10、F12、F14、F15、F16、F18；而这些传感器所获得的产品在时间维度上涵盖了从 1992 年到 2013 年所有影像。如图所示：

| Year\Sat. | F10 | F12 | F14 | F15 | F16 | F18 |
|---|---|---|---|---|---|---|
| 1992 | F101992 | ------- | ------- | ------- | ------- | ------- |
| 1993 | F101993 | ------- | ------- | ------- | ------- | ------- |
| 1994 | F101994 | F121994 | ------- | ------- | ------- | ------- |
| 1995 | ------- | F121995 | ------- | ------- | ------- | ------- |
| 1996 | ------- | F121996 | ------- | ------- | ------- | ------- |
| 1997 | ------- | F121997 | F141997 | ------- | ------- | ------- |
| 1998 | ------- | F121998 | F141998 | ------- | ------- | ------- |
| 1999 | ------- | F121999 | F141999 | ------- | ------- | ------- |
| 2000 | ------- | ------- | F142000 | F152000 | ------- | ------- |
| 2001 | ------- | ------- | F142001 | F152001 | ------- | ------- |
| 2002 | ------- | ------- | F142002 | F152002 | ------- | ------- |
| 2003 | ------- | ------- | F142003 | F152003 | ------- | ------- |
| 2004 | ------- | ------- | ------- | F152004 | F162004 | ------- |
| 2005 | ------- | ------- | ------- | F152005 | F162005 | ------- |
| 2006 | ------- | ------- | ------- | F152006 | F162006 | ------- |
| 2007 | ------- | ------- | ------- | F152007 | F162007 | ------- |
| 2008 | ------- | ------- | ------- | ------- | F162008 | ------- |
| 2009 | ------- | ------- | ------- | ------- | F162009 | ------- |
| 2010 | ------- | ------- | ------- | ------- | ------- | F182010 |
| 2011 | ------- | ------- | ------- | ------- | ------- | F182011 |
| 2012 | ------- | ------- | ------- | ------- | ------- | F182012 |
| 2013 | ------- | ------- | ------- | ------- | ------- | F182013 |

Average Visible, Stable Lights, & Cloud Free Coverages

图 1 DMSP\OLS 的稳定无云夜光遥感影像产品

由上图可以看出，DMSP/OLS 产品来源复杂，再加上一些其他因素的影响，使得此类产品存在图像的

不连续和过饱和，使得不同数据之间的可比性比较差。这就导致了不同的影像之间是不可比的，导致了研究者无法在时空域上对于遥感影像进行研究，限制了研究范围。而夜间灯光饱和部分数据监测值无法反映真实的夜间灯光现象，因此也就不可以反映真实的人类活动状况，使得研究者的研究精度降低[8]。而且同时,不同的 OLS 传感器在获取影像时并没有进行星上辐射校正，使得即使是同一传感器所获得的连续图像也存在着 DN 值异常波动[9]。因此，对于此类产品进行先行校正是非常重要的。

进行校正之后的图像，就可以在时空域进行宏观数据采集和分析了。这种预处理可以保证研究者所进行的研究具有可信性。

### 2.1.1 相互校正

为了方便进行校正，笔者期望将图像默认坐标系 WGS-84 坐标系转为兰伯特等面积投影，并且将网格重采样大小设置为 1km²。研究者可以将黑龙江省的鸡西市作为伪不变目标区域，因为此处 1992 年到 2013 年间城市发展比较稳定，而且灯光的 DN 值范围比较广。以此进行提取的数据作为校正数据集，可以对比其他的数据进行校正[10]。

经过相关学者的研究，可以发现，用二次方程对于其他图像进行拟合修正效果最好，相关系数可以达到最高。研究者利用下面的式子来进行回归拟合：

$$DN' = a \times DN_0^2 + b \times DN_0 \quad (1)$$

经过试验，每份产品所采取的系数表如下所示：

| 数据集 | $a$ | $b$ | $c$ | SSE | $R^2$ |
|---|---|---|---|---|---|
| F16→F15 | −0.001 447 | 1.091 | 0.913 | 8.531e+04 | 0.930 9 |
| F15→F14 | −0.003 202 | 1.093 | 1.766 | 1.523e+05 | 0.864 6 |
| F14→F12 | 0.003 413 | 0.628 | 2.717 | 2.337e+04 | 0.956 4 |
| F12→F10 | 0.001 906 | 0.832 | 0.886 | 1.255e+04 | 0.917 1 |
| F16→F18 | 0.004 262 | 0.673 | 0.766 | 5.901e+04 | 0.895 5 |

图 2 校正方程的系数表[11]

研究者可以采用如上的参数对于影像进行修正。笔者在此处使用 ArcGIS 的栅格计算器功能完成了栅格的校正，在此不加详叙。

不过经过观察研究者可以发现，有些传感器的产品在同一年度上具有重复产品。这两份产品从理论上来说，经过校正之后，应当是两份相同的产品。但事实上，两份产品由于传感器的性能不同而具有差异：差异一方面是来自于不稳定像元的出现，一方面稳定像元的 DN 值也并不是处处相同的。有鉴于此我们可以对于两份产品做平均处理，即：

$$DN_0 = (DN_n^a + DN_n^b)/2 \quad (2)$$

按照上面的式子我们对于影像进行再一次处理，即可以得到经过平均的图像，并且可以把该图像作为该年度的唯一图像来进行使用。

### 2.1.2 基于年份的连续性校正和饱和校正

即使进行了上面的校正之后，也并没有解决我们的 DN 值饱和问题，而这个问题会干扰研究者对于夜光遥感数据进行的分析和评估。因此做好年度的连续性校正是很有必要的。根据已有的研究，大家可以知道的是，按照我国的经济发展规律，结合 DMSP/OLS 的 Stable_Light 产品的特点[12]，存在如下的关系：

$$DN_i \geq DN_j \quad While\ i > j\ and\ i,j\ represents\ years \quad (3)$$

这样就解决了连续性问题，即研究者应当保证本年度图像的 DN 值不小于往年的图像。该解决方案借用各种软件或者其他工具依照上述表达式进行。

同样，研究者也要解决过饱和的问题，因为 DN 值的过度饱和会让研究者不能够对于区域影像进行正确的分析，使得处理成果不能够正确地反映区域的真实经济发展情况。有鉴于此，相关学者提出应当削去

遥感影像上 DN 值超过 63 的部分，即：

$$DN_n = \begin{cases} 63 & if\ DN_n > 63 \\ DN_n & if\ DN_n < 63 \end{cases} \quad (4)^{[11]}$$

按照这种原则，研究者可以将图像的过饱和区域削去，避免 DN 值过饱和引起的一些异常值的产生，使得图像可以被更好地使用。

## 2.2 影像的区域分割

### 2.2.1 区域分割的分析

研究者希求对于影像进行区域分割，然后根据分割的区域，对于每个分块内的 DN 值进行各种统计，从而获得我们所需要的数据。那么，对于中国区域进行简单而合理的划分是十分重要的。由于较小的岛屿、海洋的夜光占比是非常小的，而且也并不是城市或者城市区域的主要存在地区，因此本次区域分割将不包括对于海洋的分割和对较小岛屿的分割(如中国的南海诸岛，而是面向中国大陆和中国台湾岛以及其周边距其较近的岛屿。

相关学者就经济发展情况对于中国大陆地区进行东、中、西三带划分。实际上划分方法是多种多样的。根据鲁凤、徐建华 (2004) 指出，研究者可以使用两分法、三分法、四分法、六分法、七分法、省级行政单位划分以及县级行政单位划分来对于自己的研究区域进行分割。而对于三分法，即将中国划分为东部、中部、西部，则是经济研究者常用的一种手段。相关研究者对于西、中、东三地区进行经济分析，得出了三地区的经济发展的不同之处[13]。

不过笔者在研究中发现这样的划分并不是非常细致，即这样的划分不能满足笔者的研究需要。比如，同样是西部地带，青藏地区和西北地区是不同的：首先青藏地区和西北地区的地质不同，造成了两个地区的经济增长和居住情况的差异(譬如，虽然青藏地区和西北地区一样都有大片的不宜居区域，但是很明显，西北地区有比西藏更多的宜居区域和沙漠绿洲，而且对于沙漠和戈壁地区的改造显然比在高原和山峰地区开拓居民点容易得多)；而且，青藏地区的交通环境相比西北地区来说更加恶劣；同时，青藏地区的农业规模相比西北地区来说也是很小的，因为西北地区的干旱地带也有着很多种农作物耕地，更不用提新疆有些地区甚至可以生产水稻[14]。这种种差异表明将一些地区笼统划分到一起并不是很严谨的。

当然，在省级行政区划分中，研究者可以看到省份内部的产业差异和经济发展差异也是存在的；但这并不是很严重的问题，因为省级行政区内各方面的差异比较小。较为复杂的地区，类似于云南、四川地区，其处于第一、二地理阶梯交界处，地势复杂，地质组成繁多，并且拥有很多不同的经济类型。比如，四川地区的畜牧业就有着地域上的不同分布和不同发展[15]。

这样就造成了一个矛盾，即较大范围的区域研究总是面临着区域内情况有着明显不协调的问题。那么为了避免这一点，研究者习惯于将研究范围细致化。但是细致化就会带来一些问题，即大量小区域的分析和聚类会使得处理过程变得非常繁琐，而且小区域的分别研究也不能呈现出大区域的时空宏观性。因此研究者们希望对于区域的划分能够保持一个恰当的量级，而且划分的方式可以尽量简便。

在这里，有学者对于中国各省份发展潜力做了一些研究，使用区间 DBSCAN 算法对各个省份发展潜力做出评估，得到如下示意图：

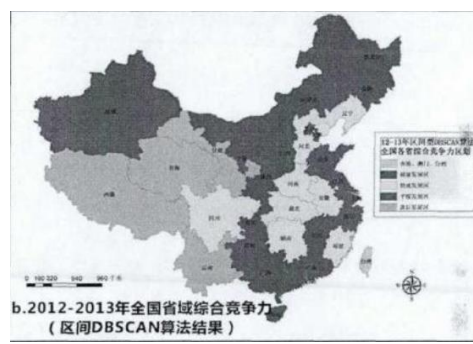

图 3 2012-2013 年全国省域综合竞争力[16]

由此可以看出，按照这样的分类方法具有一定的合理性，但是对于本研究而言，这种区域划分无疑是割裂了区域的统一性和一致性，使得全国各大地区出现了空间域上的断裂。

2.2.2 区域分割的结果

由上述的一些分析，研究者若要选出一种合适的区域划分，就应当兼顾区域的一致性、区域的时空域联合性以及区域发展一致性。划分的时候，同一个区域内的差异应当尽可能少，而且这样的区域划分应当考虑到操作的简便性和数据分析的难易度。研究者以上述的一些事实和划分方法作为框架，施行了八划分法，中国大陆和中国台湾划分为八个区域：西北地区、西南地区、青藏地区、华北地区、华中地区、东北地区、华东地区、华南地区。具体划分如图所示：

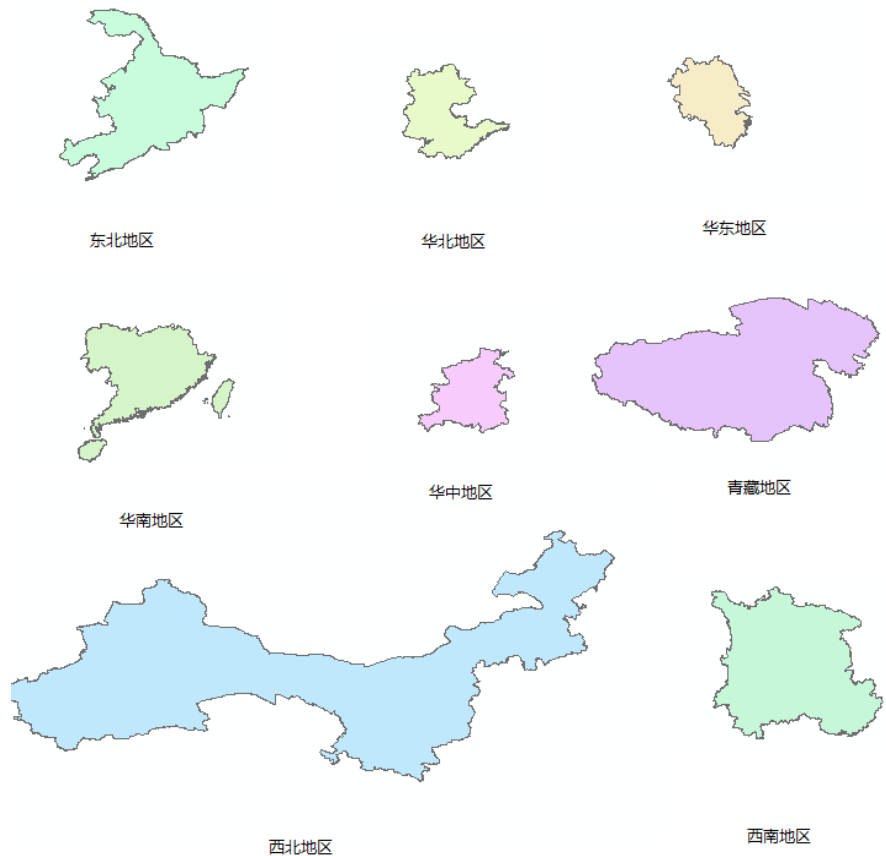

图 4 区域划分示意图

研究者可以按照这个模板对于处理好的夜光遥感影像进行裁剪，进而统计分析其 DN 值数据。其中华中地区包含河南省和湖北省，覆盖省份较小，是可以被归并为其他区域的，但是河南的众多人口和湖北省的武汉地区城市群的建设使得其被单独列出，以进行进一步的处理和研究。西北地区占地广泛，包括内蒙古、山西、陕西、新疆、甘肃、宁夏。这样的考虑明显是按照传统的西北地区的概念进行划分的；而且陕西的西安地区明显是西北重镇，考虑到此对于西北地区经济发展所进行的带动作用，把秦、晋地区纳入西北地区范围内也是有必要的。

以上为区域的划分问题的讨论。

## 2.3 **数据的提取**

在进行数据采集的时候，研究者期望保持底图的 Lambert 等面积投影，进而可以在与现实面积比相同

的情况下，对于各区域的所采集的数据进行相互比较。所采集的数据包含像元量、区域面积、DN 最大值、DN 最小值、DN 值区间长度、DN 值均值、DN 值标准差以及区域内 DN 值总和。在进行区域划分的前提下，研究者将把这些数据从 1992 年~2013 年这 22 年的夜光遥感影像中提取出来。然后，研究者把这些数据进行整合，将每个区域的 1992 年~2013 年的数据整合在一起，且把各个年份的 8 个区域的数据整合在一起，方便研究者在时空域上研究这些资料。到最后，可以得到总共 30 份(22+8)份基础数据文件。

关于地域分区的 8 份文件，研究者可以针对各个地域的年度发展进行数据分析，而关于年份的 22 份数据，研究者则可以平行比较各个地区的发展差异和这些差异在各个年份的变动。基于这样的基础数据提取和聚类，研究者可以在时间和空间上对于夜光遥感影像所反映的情况进行比较探究。

# 3 区域发展的年度分析

## 3.1 各区域年度拟合方程和回归模型的建立

上面提取的数据的重点内容明显是 DN 值均值和 DN 值标准差。研究者对于所提取的各区域年度数据的 DN 值均值和标准差进行折线图展示，如下图所示：

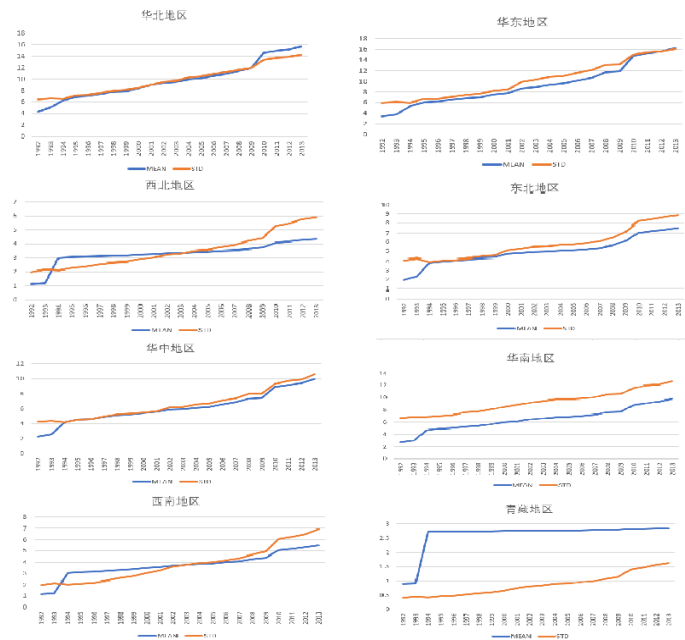

图 5 各区域年度夜光 DN 值发展曲线

按照这样的折线图，可以发现这些发展曲线是适合进行一次拟合(线性)或者二次拟合的。其中研究者对于青藏地区的 DN 值平均值发生跃迁事件进行了分析，发现 1993 年~1994 年的图像 DN 值最小值发生了突变，即由 0.886 跃迁至 2.717。青藏地区人烟稀少，城市区域零散且分布较少，因此倘若图像的 DN 值最小值由于处理原因发生跃迁，则其 DN 值平均值将受到较大影响。事实上，可以看出其他地区都在 1993~1994 时段发生了或多或少的跃迁。为了保持一致性，故对于回归模型的建立则采取 1994 年及其之后的数据。在对于数据进行筛选过后，研究人员对于原始提取数据进行一系列的回归分析(以[年份]-1993 为 X，以均值为 Y)。

| 地区 | 系数 a | 系数 b | $R^2$ | F 值 |
| --- | --- | --- | --- | --- |
| 东北地区 | 4.06574 | -0.01932 | 0.962576 | 218.6281 |
| 华北地区 | 6.711705 | 0.076032 | 0.974729 | 327.8572 |
| 华南地区 | 4.758388 | 0.086011 | 0.981561 | 452.4728 |
| 华中地区 | 4.507647 | 0.013002 | 0.978565 | 388.0485 |
| 青藏地区 | 2.734711 | -0.00129 | 0.981735 | 456.8688 |
| 西北地区 | 3.090963 | -0.01608 | 0.975293 | 335.5252 |
| 西南地区 | 3.139074 | -0.00641 | 0.97894 | 395.1068 |
| 华东地区 | 5.782286 | 0.047853 | 0.98293 | 489.4565 |

表 1 线性回归分析

这个结果表明，对于数据进行二次拟合处理，可以得到很高的 R² 值，其值基本可以达到 0.96 或者以上；并同时具备很高的 F 值以及非常低的对应的 P 值，说明该拟合模型是有效的。

所以可以得出该结论：中国各区域的 DN 值均值增长进程是可以被认为是符合线性增长的。研究者将使用这些分析数据进行下一步的研究。

## 3.2 区域发展的速度分析

根据上面的回归处理，研究者选择了线性回归模型作为区域经济研究模型。但是 DN 值并不一定是和真实的经济状况呈现线性对应状态的，对于此，研究者还应当进行进一步说明，阐明 GDP 和 DN 值的具体关系。在此之后，研究者可以利用 DN 值进行对应的区域发展分析。

### 3.2.1 区域发展与 DN 值关系

在进行无月光影响的遥感探测时，研究人员发现人类集群区域的信号响应是最为强烈的，而一些其他的辐射源，诸如山火等，在夜光遥感影像中都是以噪声形式存在[17]，经过平滑后很容易被削去，亦或者是其影响在进行一定范围的数据统计时被其他的数据弱化。所以夜光遥感影像上所获取的 DN 值是具有很高的参考价值的。

为了方便进行与现实数据的比对，这里将引入 TNL 概念，即区域内灯光像元不同等级亮度值与对应像元数的乘积之和：

$$TNL = \sum_{i=1}^{C} DN_i \quad (5)$$

在采用采用线性、指数、二次项、乘幂 4 种回归模型对 DMSP 夜间灯光与 GDP 的空间关系进行分析之后，相关研究者得到 4 种模型相关系数分别为 0.7817、0.7834、0.8016、0.8077。由此可见这 4 种模型的差异并不是很大，而且契合度都是比较高的[18]。其中显而易见，乘幂模型是最优模型。根据文献[18]所阐述，在校正后的 DMSP 影像之中，乘幂模型所拟合的 GDP 发展状况和真实情况的相关系数是更高的，使得 TNL 和真实 GDP 之间拥有了很高的相关度。因此，是可以使用乘幂模型进行区域发展分析的。

| 数学模型 | 数学表达式 |
|---|---|
| 线性 | yGDP=ax+b |
| 指数 | yGDP=ae$^x$ |
| 二次项 | yGDP=ax$^2$+bx+c |
| 乘幂 | yGDP=ax$^b$ |

表 2 几种数学模型

很明显，在这里研究者所使用的 DN 值均值即：

$$Mean = \frac{TNL_n}{COUNT} \quad (6)$$

对于各区域而言，值 Square 是恒定不变的(在这里 Square 可以指的是区域内像元总量)，因此 Mean 值是和 TNL 值保持线性关系的。既然存在这样的连锁的线性关系，所以研究人员是可以将 DN 值均值的乘幂结果代表区域内经济发展水平的。因此在下文中，研究人员将采用区域发展指数(Regional Development Index, RDI)来代替原先的 DN 值称呼。

### 3.2.2 基于线性模型的区域发展速度与 GDP 增速分析

系数 a 即线性回归模型的一次项的参数。通过对于函数进行一次求导可得到系数 a 作为发展增长速度。下图则显示出土地平均 RDI(下文皆称为 Land Average RDI, LaRDI)的增长速率：

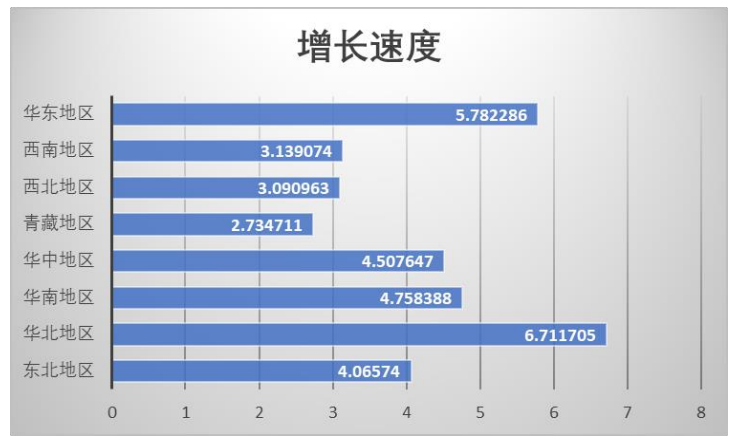

图 6 LaRDI 增长速率

在这里可以看出，华北地区是地均发展最为强劲的地区，紧随其后的是华东地区。而青藏地区则处于倒数，同时西南、西北地区的发展速率也是相对较低的。现在研究者将从国家统计局的年度 GDP 统计数据中，对于各省份的年度 GDP 增长进行可视化展示，得到如下效果(选取数个省份进行分析)：

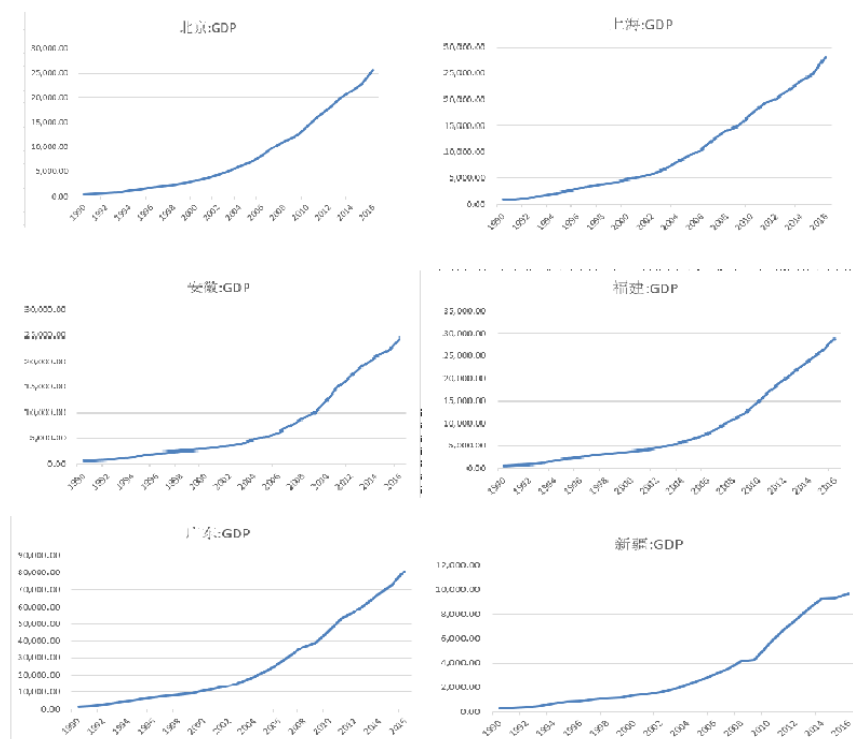

图 7 各省 1990 年到 2016 年的 GDP 产值折线图

按照折线图，可以看出发展曲线是倾向于用二次回归模型进行模拟的。研究人员在对于这些数据进行回归分析的时候，分别进行了一次回归分析和二次回归分析，发现二次回归是更加贴合省份发展情况的，其相关系数达到 0.97~0.98。

经过对比可知，一次回归模型和二次回归模型都有着很高的相关系数，但是二次回归模型明显是与实际数据有着很高的契合度的，因此可以选择采用二次回归模型来模拟 GDP 发展函数。其中需要说明的是，DN 值模拟模型是以[年份]-1993 作为自变量 X 的，DN 值均值(或者是 LaRDI)为 Y；而此处则是使用实际年份作为自变量 X，以实际 GDP 产值作为 Y。

现在研究者进行区域整合，使得数据归化为八块既定区域的数据，然后再次进行区域数据增长的拟合分析。

现在将省份 GDP 数据按照既定划定范围进行求和，即将这些省份的 GDP 分类、合并到之前已经设定的八个区域内。按照之前的 LaRDI 的计算标准，此处使用 DN 像元数量作为分母，对 GDP 的量进行平均，以方便计算。不同省份的数据合并将按照下面的原则进行：

$$AGDP_n = \frac{\sum GDP_n^i}{\sum COUNT_i} \qquad (7)$$

在进行聚合之后，抽取样例观察其发展折线图，得到下面的结果：

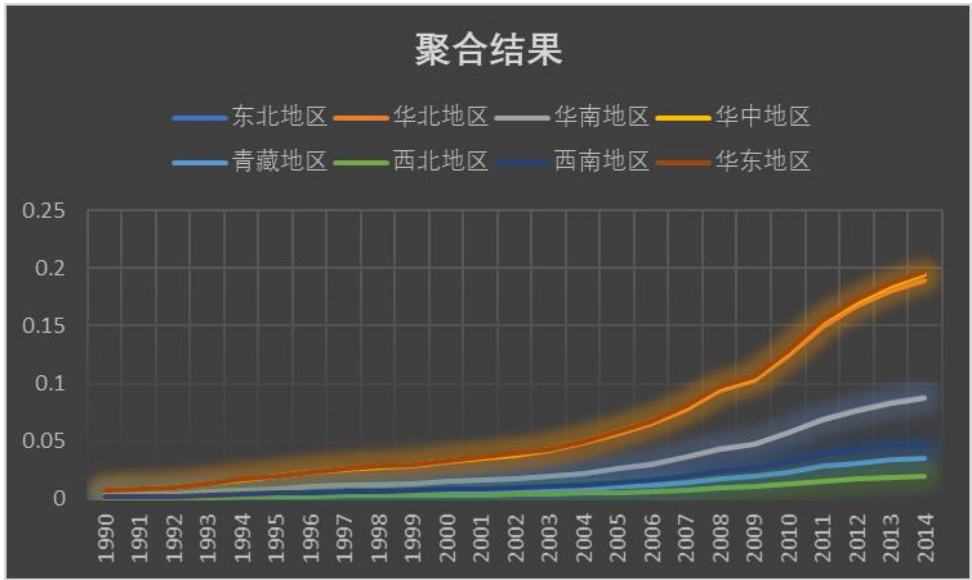

图 8 聚合后各地区 GDP 增长曲线

尝试进行回归模型分析，将([年份]-1989)/100 作为自变量 X，进行幂函数拟合，得到如下图所示的结果(以东北地区为例)：

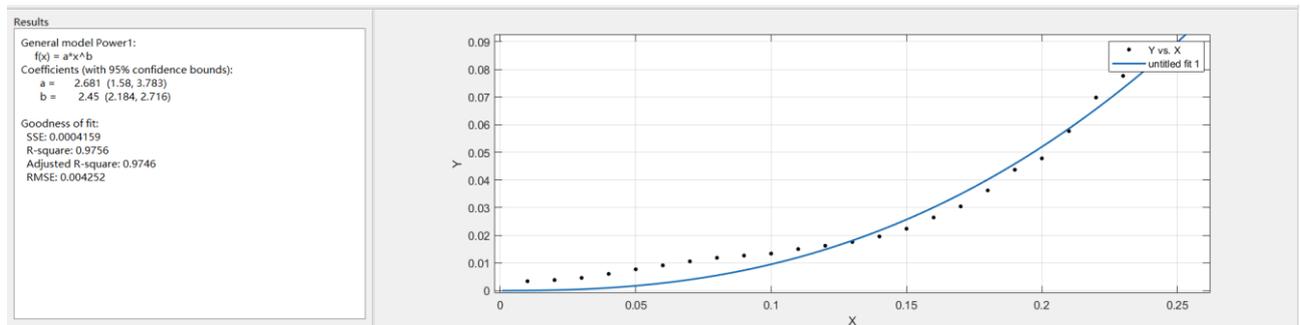

图 9 幂函数拟合后的结果

由此可见，其相关系数达到很高的值。笔者研究了其他几个地区的幂乘函数模型，发现相关系数都非常高。结合原来的 LaRDI 的数据进行比对,可以发现线性的 LaRDI 增长模型是可以由文献[18]所提到的 TNL-GDP 幂乘函数模型，对应幂乘函数的 GDP 增长模型的。在这里，研究者发现 LaRDI 模型可以在宏观的时间段内给出一个大致平稳的发展速度，弱化了幂乘函数模型的不稳定的增长型速度的影响，方便人们对于各个地区发展状况有个宏观的认识，从而对于各地区的本质上的发展问题进行研究。

## 3.3 区域发展的不均衡问题

### 3.3.1 标准差模型的拟合分析

对于数据中的 STD(DN 值标准差项)进行分析也是很重要的，STD 是提取数据中的重要组成部分。现在对于数据进行回归分析,研究人员决定采用 Fourier 模型进行数据拟合,得到结果如下(为东北地区的 Fourier 回归拟合模型)：

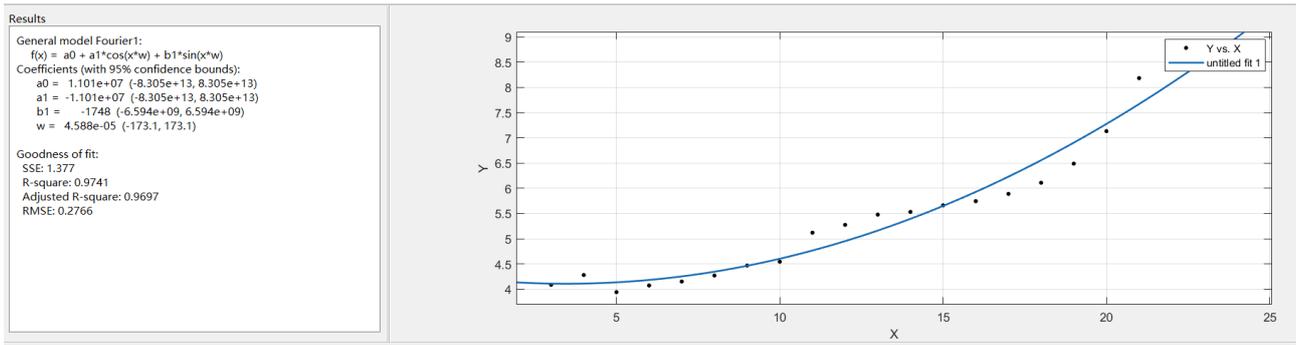

图 10 Fourier 回归模型(东北地区)

可以看出，Fourier 模型可以使得回归系数达到 0.96~0.99 左右，是采用的几种模型中最好的一组模型。Fourier 模型呈现如下式：

$$y_x = a_0 + a_1 * \cos(\omega * x) + b_1 * \sin(\omega * x) \quad (8)$$

在对于数据进行拟合过后，观察其曲线和曲线函数，可以感受到 STD 增速是持续升高的。现计算各个地区的 STD 从 1992 年~2013 年平均增速，得到如下图所示的结果：

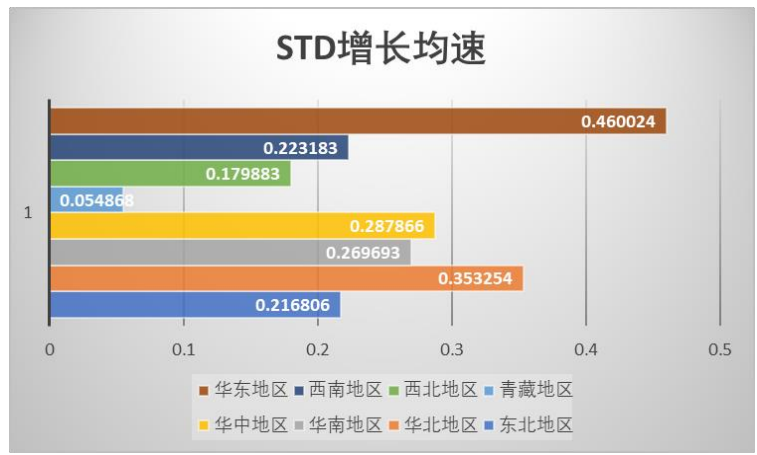

图 11 STD 增长均速

基于此，研究者可以发现青藏地区的 STD 增长最为缓慢，而华东地区的 STD 增长则是处于明显的巅峰地位。

3.3.2 STD 增长模型反映的地区发展问题

在 LaRDI 模型中，地区的 DN 值总量(即 TNL 指数)在某种程度上代表了一个地区的经济总量，那么 STD 量则代表了地区发展的不均衡度。一个地区的 STD 的值越大，则代表了一个地区的发展越不均衡。这种不均衡并不一定能被解释为某种负面的信号。事实行，我国不同的区域始终存在一定的经济差距。改革开放以来，我国经济态势迅猛增长，GDP 发展达到了一个前所未有的水平和速度，但也正是随着各种发展，导致了区域发展不平衡、不协调因素的存在[19]。

从宏观态势上，研究者们可以发现中国主要的发展不协调的矛盾存在于东部沿海和中西部内地区域发展不平衡、城乡发展不平衡以及发达地区和欠发达地区不平衡等现象当中[20]。但从划区域的角度来说，不平衡的现象则以一种全新的态势展现出来。

首先，经济增长会带来发展不平衡的存在和加剧。由图 11 可以看出，所有地区的 STD 增长均为正值，而这些地区都处于经济增长的状态。与实际情况进行比较可以发现，中国的经济发展是"先富带动后富"的方式，各个地区内一些得天独厚的子区域因为受到一些政策的影响而首先崛起，造成了一批较高的 DN 值像元出现，而且其余的很多像元仍然维持原先低 DN 值状态，拉大了各个像元 DN 值的间距，造成 STD 值的连续增长[21]。

其次，各个地区的 STD 和 LaRDI 增速排名相比大致相当。这表明地区经济增长越快，就会导致地区发

展的越不均衡。这种现象在华东地区得到了极为明显的体现，同样也出现在经济增速较高的华北地区等区域。而发展缓慢的青藏地区的 STD 发展速度是最低的。初步解释是：沿海发达地区的经济增长速度极快，与和其为同一区域的地域产生了很大的发展间壑，导致了区域发展的极为不均匀，是一种由于发展态势迅猛导致的 STD 增长过快现象。另一方面，内地的很多区域受到沿海发达区域的影响较小，经济发展较为平缓，所以没有产生此类现象。

# 4 城市化水平评估

在此处，研究者将利用所提取的数据对于各个区域的城市化进展进行探究，以体现上世纪 90 年代以来我国如火如荼的城市化进程。研究者在此选择华北地区、华中地区、华东地区这些沿海发达地区作为研究对象，并选择 1994 年、2000 年、2005 年、2013 年为时间节点，对于这些研究对象进行时间上的发展分析。

参照某些学者的理论，在这里研究者将 DN 值 55 作为大城市的 DN 值阈值[22]。较高的 DN 值既可以大致提取出城市的繁华区域的核心，也可以尽力避免一些 DN 值较高的噪点对于结果所产生的影响。经过相关统计，可以对于这三个地区的数据进行制表如下：

| 华北地区城市化的 DN 数据 | |
| --- | --- |
| 年份 | DN 值>=55 像元数量 |
| 1994 | 1063 |
| 2000 | 2032 |
| 2005 | 4272 |
| 2013 | 12933 |

表 3 华北地区城市化

| 华东地区城市化的 DN 数据 | |
| --- | --- |
| 年份 | DN 值>=55 像元数量 |
| 1994 | 790 |
| 2000 | 1230 |
| 2005 | 3008 |
| 2013 | 15309 |

表 4 华东地区城市化

| 华中地区城市化的 DN 数据 | |
| --- | --- |
| 年份 | DN 值>=55 像元数量 |
| 1994 | 182 |
| 2000 | 274 |
| 2005 | 625 |
| 2013 | 3310 |

表 5 华中地区城市化

首先研究者对于这些数据进行横向对比，发现华中地区的大城市群的量级一直是最低的：这是可以理解的，因为华中地区仅包含湖北和河南两省，这两个省份并非处于沿海经济地区。就湖北省而言，大武汉经济圈囊括了湖北省最主要的城市化区域，使得城市规模位居第二的荆州的 GDP 也仅仅为武汉市的 1/7。武汉市 2000 年的国内生产总值达到 1206.84 亿元，占全省的 28.2%，体现出了湖北省的经济集中型的城市建设方法[23]，也使得湖北省其他地区不能尽快地被武汉地区的发展所带动，导致了华中地区的繁华城市群的集中化。

通过对于原先遥感影像进行二值化处理(阈值为 55)，研究者可以发现，几个阶段以来，华中地区始终以郑州城市群、武汉城市群为核心进行发展，从 1994 年~2005 年，这二者之外的大城市群是没有明显出

现的。但是到了 2013 年，原先的一些不明显的城市地区基于之前的城市地区有了明显的拓张，其中河南省的这一特点的更加显著的，全省各个地区都有了星罗棋布的繁华城市点群，并有了一定的互相融合的趋势；而湖北省则坚持大武汉战略，使得武汉称为该地区最为强大的城市群，但是省内其余地区没有出现太多的新生繁华城市区。

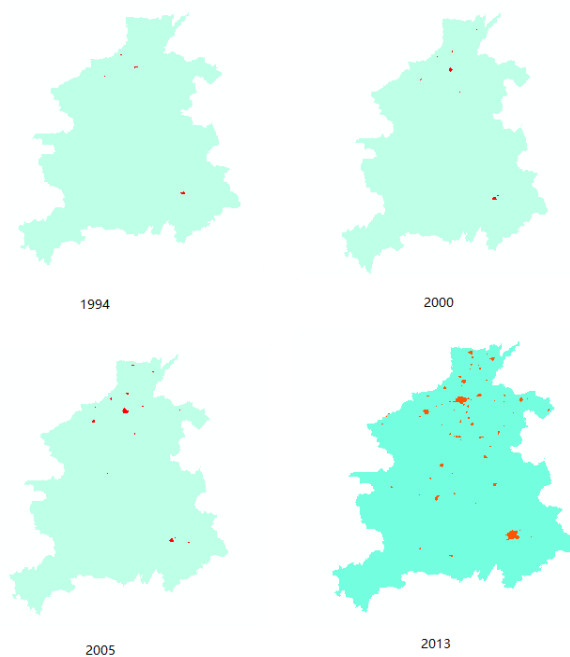

图 12 华中地区城市群发展

至于华东地区，研究者可以发现该地区的繁华城市群规模一开始一直低于华北地区，但是在 2013 年附近反超华北地区，拥有了无与伦比的超大的城市集群。一方面，华东地区很早就拥有了上海大城市群作为底子，另一方面，华东地区靠近海岸，拥有得天独厚的经济潜力，因此发展一向迅猛。华东地区囊括我国经济最为发达的长三角地区，具有强大的交通网络，足以引领全国经济的发展[24]。与此同时，相较于内地而言，由于地区发展足够快，经济足够发达，华东地区也会产生明显的溢出效应，使得繁荣城市群的扩张变快，具有很明显的城市吞并效应。

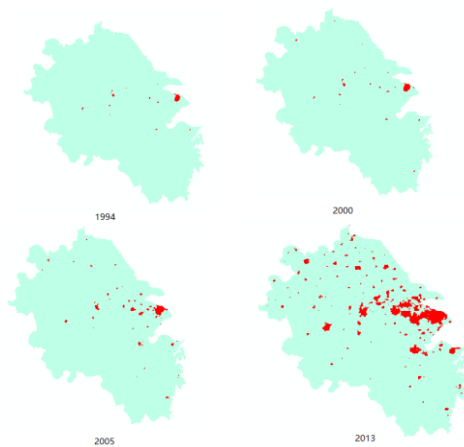

图 13 华东地区城市群发展

如图所示，1994 年华东的基础城市群主要是上海城市群，而到了 2013 年，上海城市群的规模急速扩大，并且与苏州城市群等其他城市群发生了交融、吞并现象。与此同时南京、合肥、杭州、温州城市群也在急速扩大，并利用了零散的城市繁华区域(多为中小型城市)，实现了区域的扩大，并可能在未来实现区域的衔接，形成超级城市群。

华北地区的繁华城市群规模是以北京为基础的。北京市是中国的首都，是最初的大规模聚落，这使得

华北城市群有着先天优势。但是受到政策、地理因素等影响,其未能如华东地区那样保持很高的发展速度,最终被华东地区超越。

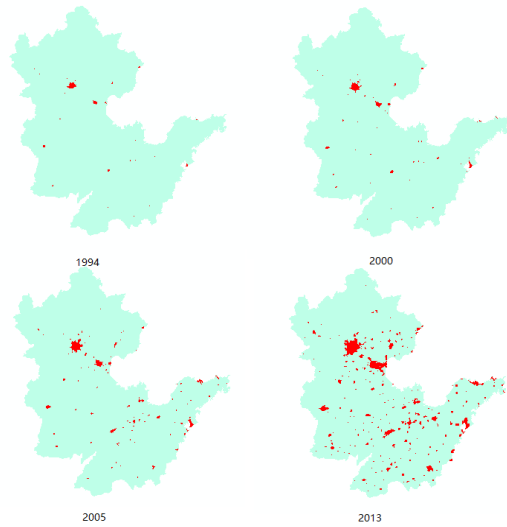

图 14 华北地区城市群发展

但这并不代表华北城市群的衰败。恰恰相反,由图 14 可知,华北地区原先只有北京城市区作为大规模城市区,而到了 2013 年,华北地区拥有了北京、天津、青岛、石家庄、济南等多核心城市集群。其中,天津和北京城市群是最大的,并且有很强的衔接趋向。而且山东和河北的城市群星罗棋布,多方位发展,到最后可能建设成一个强大的连续城市集群[25]。

# 5 结论

对于夜光遥感的分析比进行周期性的传统普查具有更高的效益。基于这样的原则,研究者基于 DMSP/OLS 产品,对于区域发展和城市化进程进行定量和定性分析,综合研究了 1994 年~2013 年间中国的发展趋势,得出了相关的数据和结论:

(1)对于 DMSP/OLS 产品进行预先校正是必要的,这样可以大大提高 TNL 与实际 GDP 的拟合精度;
(2)使用 LaRDI 模型可以给出一个宏观范围上的大致发展速度,便于研究人员进行运算和分析;
(3)经济发展会带来一些发展不协调、不平衡问题,而且此类问题在未来会越来越显著,因此应当尽量做好区域发展的协调;
(4)区域发展带来的城市化趋势是愈演愈烈的,并且会导致城市的吞并效应,形成较大规模的连续城市集群,而这可能是以后城市发展的主导方向。

# 6 参考文献